\documentclass[pra,twocolumn,showpacs,amsmath,amssymb,superscriptaddress]{revtex4}

\usepackage{graphicx}
\usepackage{mymacros}
\newcommand{\olamb}{\overline{\lambda}}

\newcommand{\void}[1]{}
\renewcommand{\nu}{2\epsilon-\Delta}

\begin{document}

\title{Qubit-oscillator dynamics in the dispersive regime:
\\ analytical theory beyond rotating-wave approximation}

\author{David Zueco}
\affiliation{Institut f\"ur Physik, Universit\"at Augsburg,
        Universit\"atsstra{\ss}e~1, D-86135 Augsburg, Germany}
\author{Georg M. Reuther}
\affiliation{Institut f\"ur Physik, Universit\"at Augsburg,
        Universit\"atsstra{\ss}e~1, D-86135 Augsburg, Germany}
\author{Sigmund Kohler}
\affiliation{Instituto de Ciencia de Materiales de Madrid, CSIC,
	Cantoblanco, E-28049 Madrid, Spain}
\author{Peter H\"anggi}
\affiliation{Institut f\"ur Physik, Universit\"at Augsburg,
        Universit\"atsstra{\ss}e~1, D-86135 Augsburg, Germany}

\date{\today}

\begin{abstract}
We generalize the dispersive theory of the Jaynes-Cummings model
beyond the frequently employed rotating-wave approximation (RWA) in
the coupling between the two-level system and the resonator. For a
detuning sufficiently larger than the qubit-oscillator coupling, we
diagonalize the non-RWA Hamiltonian and discuss the differences to the
known RWA results.  Our results extend the regime in which dispersive
qubit readout is possible.  If several qubits are coupled to one resonator,
an effective qubit-qubit interaction of Ising type emerges, whereas RWA
leads to isotropic interaction.  This impacts on the entanglement
characteristics of the qubits.
\end{abstract}

\maketitle


\section{INTRODUCTION}
More than forty years ago, Jaynes and Cummings \cite{Jaynes1963}
introduced a fully quantum mechanical model for the interaction of
light and matter, which are represented by a single harmonic
oscillator and a two-level system, respectively.  Within dipole
approximation for the interaction, that model is expressed by the
Hamiltonian    
\begin{equation}
\label{HRabi}
 H =  \frac{\hbar\epsilon}{2} \sigma^z + \hbar \omega a^\dagger a +
 \hbar g  \sigma^x (a^\dagger +a) ,
\end{equation}
where $\hbar \epsilon$ is the level splitting of the two-level system,
henceforth ``qubit'',
$\omega$ is the frequency of the electromagnetic field mode, and $g$
the dipole interaction strength. The Pauli matrices $\sigma^\alpha$,
$\alpha = x,y,z$, refer to the two-level system, while $a^\dagger$ and
$a$ denote the bosonic creation and annihilation operators of the
electromagnetic field mode.  This
model describes a wealth of physical phenomena rather well and by now
is a ``standard model'' of quantum optics.  A particular experimental
realization  of the Hamiltonian is an atom interacting with the field
inside an optical cavity, usually referred to as cavity
quantum electrodynamics (cavity QED).  Corresponding experiments have
demonstrated quantum coherence between light and matter manifest in
phenomena such as Rabi oscillations and entanglement
\cite{Raimond2001a, Walther2006a}.  

Despite its simplicity, the Hamiltonian \eqref{HRabi} cannot be
diagonalized exactly and, thus, is often simplified by a rotating-wave
approximation (RWA).  There, one expresses the qubit-cavity
interaction in terms of the ladder operators $\sigma^\pm =
\frac{1}{2}(\sigma^x \pm i\sigma^y)$.  In the interaction picture with
respect to the uncoupled Hamiltonian, the coupling operators  
$\sigma^+ a$, $\sigma^- a^\dagger$ and $\sigma^- a$, $\sigma^+
a^\dagger$ oscillate with the phase factors $\exp [\pm \iu (\omega -
\epsilon)t]$ and $\exp [\pm \iu (\omega + \epsilon)t]$, respectively.
Operating at or near resonance, the cavity-qubit detuning is
small, $\left|\epsilon - \omega \right| {\ll}\, \epsilon + \omega$, so
that the former operators oscillate slowly, whereas the latter exhibit
fast ``counter-rotating'' oscillations.  If in addition, the coupling
is sufficiently weak, $g \, {\ll} \min \{\epsilon, \omega\}$, one can
separate time scales and replace the counter-rotating terms by their
vanishing time average.  Then one obtains the Jaynes-Cummings
Hamiltonian \cite{Jaynes1963}
\begin{equation} \label{HJC} H_\mathrm{RWA}
= \frac{\hbar\epsilon}{2} \sigma^z + \hbar\omega a^\dagger a + \hbar g
(\sigma^-a^\dagger   + \sigma^+a) \; .  \end{equation}

Lately, new interest in Jaynes-Cumming physics has emerged in the
solid state realm. There, one implements artificial atoms
with Cooper-pair boxes (charge qubits) \cite{Wallraff2004a} or
superconducting loops (flux qubits) \cite{Chiorescu2004a}.  The  role
of the cavity is played now by a transmission line or a SQUID
depending on the architecture \cite{Blais2004a, Deppe2008a}, or even a
nanomechanical oscillator \cite{Lahaye2009a}. Since the first
experimental realizations in 2004 \cite{Wallraff2004a,
  Chiorescu2004a}, a plethora of results has been obtained, such as
quantum-non-demolition-like readout of  a qubit state
\cite{Lupascu2007a}, the generation of Fock states
\cite{Hofheinz2008a}, the observation of Berry phases
\cite{Leek2007a}, multi-photon resonances \cite{Deppe2008a},
entanglement between two qubits inside one cavity \cite{Majer2007a,
  Sillanpaa2007a}, and the demonstration of a two-qubit algorithm
\cite{dicarlo2009}. 

These experiments have in common that they operate in the strong
coupling limit, that is, the coupling $g$ is larger than the linewidth
of the resonator.  On the other hand, $g$ is typically two orders of
magnitude less than the qubit and resonator frequencies. In this
scenario the Jaynes-Cummings model \eqref{HJC} has been shown to
describe the experiments faithfully.

Of practical interest is the dispersive limit, in which the qubit and
the resonator are far detuned compared to the coupling strength,
$g \,{\ll} |\epsilon \,{-}\,\omega|$ \cite{Blais2004a, Liu2005a}. In this
regime a non-demolition type measurement of the qubit can be performed
by probing the resonator \cite{Wallraff2004a, Lupascu2006a}.
Moreover, it is possible to simulate quantum spin chains with two or
more qubits that are dispersively coupled to one resonator
\cite{Majer2007a, dicarlo2009, Helmer2009a}.  The complementary
architecture of two cavities dispersively coupled to one qubit allows
building a quantum switch \cite{Mariantoni2008a}.
All these ideas have been developed from the RWA model \eqref{HJC} in
the dispersive limit or from according generalizations to many qubits
or many oscillators.  Thus, these theories are restricted to the range
\begin{equation}
\label{dispersiveRWA}
g \ll |\epsilon-\omega| \ll \epsilon+\omega ,
\end{equation}
where the first inequality refers to the dispersive limit, while the
second one has been used to derive the RWA Hamiltonian \eqref{HJC}
from the original model \eqref{HRabi}.

In recent experiments, efforts are made to reach an even stronger 
qubit-cavity coupling $g$.  Thus, it will eventually be no longer
possible to fulfill both inequalities \eqref{dispersiveRWA}
\cite{Bourassa2009a}.  In particular, when trying to operate 
in the dispersive limit, the second inequality may be
violated, so that RWA is no longer applicable.  Non-RWA effects of
the model Hamiltonian~\eqref{HRabi} have already been studied in the
complementary adiabatic limits $\epsilon \,{\ll}\, \omega$ and $\omega 
\, {\ll}\, \epsilon$ \cite{Irish2003a, Irish2005a,
  Johansson2006a}. Furthermore, Van Vleck perturbation theory has been
used in the resonant and close-to-resonant cases \cite{Hausinger2008a}.
Finally, polaron transformation \cite{Neu1996a}, cluster methods
\cite{Bishop2001a}, wave-packet approach \cite{Larson2007a} and even
generalized RWA approximations \cite{Irish2007a} have been considered.

Motivated by the importance of the dispersive regime and in view of
the experimental tendency towards stronger qubit-oscillator coupling,
we present  in this work a dispersive theory beyond RWA, so that the
second condition in Eq.~\eqref{dispersiveRWA} can be dropped. This
means that our approach is valid under the less stringent condition  
\begin{equation}
\label{dispersiveNonRWA}
g \ll |\epsilon-\omega| \;, 
\end{equation}
which implies that the detuning is not necessarily smaller than
$\epsilon$ and $\omega$.  In order to
set the stage, we briefly review in Sec.~\ref{sec:disp-JC} the
dispersive theory within RWA.  In Sec.~\ref{sec:dispn}, we derive a
dispersive theory for Hamiltonian \eqref{HRabi} beyond RWA, 
which we generalize in Sec.~\ref{sec:multi-qb} to the presence of
several qubits.


\section{Dispersive theory within RWA}
\label{sec:disp-JC}

The dispersive limit is characterized by a large detuning 
$\Delta \,{=}\, \epsilon-\omega$ as compared to the qubit-oscillator 
coupling $g$. Thus,
\begin{equation}
\label{dispcond}
 \lambda = \frac{g}{\Delta}
\end{equation}
represents a small parameter, while the RWA Hamiltonian \eqref{HJC}
is valid for $|\epsilon-\omega| \ll \epsilon+\omega$.
Then it is convenient to separate the coupling term from the RWA
Hamiltonian, i.e., to write $ H_\mathrm{RWA} = H_0 + \hbar g X_+ $
with the contributions
\begin{align}
\label{H0Xpm}
H_0 =& \frac{\hbar\epsilon}{2} \sigma^z + \hbar \omega a^\dagger a \;,
\\
X_\pm =& \sigma^- a^\dagger \pm \sigma^+ a \, .
\end{align}
Applying the unitary transformation,
\begin{equation}
\label{transfone}
D_{\rm RWA}= e^{\lambda X_-}
\end{equation}
\void{
and using the commutation relations
\begin{align}
\label{crrwa}
[H_0, X_-] ={}& -\Delta X_+ \;,
\\
[X_+, X_-] ={}& \sigma_z (2 a^\dagger a + 1) + \frac{1}{2}
\; ,
\end{align}
}
one obtains for the transformed Hamiltonian
$
H_{\rm disp} =
 D_\mathrm{RWA}^\dagger H_\mathrm{RWA}^{\vphantom{\dagger}}
 D_\mathrm{RWA}^{\vphantom{\dagger}} 
$
to second order in $\lambda$: $H_{\rm disp} =
 H_\mathrm{RWA}
+ \lambda [H_\mathrm{RWA}, X_-] + \frac{1}{2}\lambda^2 [[H_\mathrm{RWA}, X_-],X_-]$, which can be
evaluated to read
%
\begin{equation}
\label{HJCdisp}
H_{\rm RWA,disp} =
\frac{\hbar\epsilon}{2} \sigma^z
+ \frac{\hbar g^2}{2 \Delta} \sigma^z
+ \left(\hbar\omega 
+ \frac{\hbar g^2}{\Delta} \sigma^z \right) a^\dagger a
\; .
\end{equation} 
The physical interpretation of (\ref{HJCdisp}) is that the oscillator 
frequency is shifted as
\begin{equation}
\label{shiftRWA}
 \omega
\to \omega \pm g^2/\Delta \; , \end{equation}
where the sign depends on the state of the qubit.  If one now probes
the resonator with a microwave signal at its bare resonance frequency
$\omega$, the phase of the reflected signal possesses a shift that
depends on the qubit state.  This allows one to measure the
low-frequency dynamics of the qubit \cite{Blais2004a}.  Since,
according to Eq.~\eqref{HJCdisp}, the qubit Hamiltonian
$(\hbar\epsilon/2)\sigma_z$ commutes with the dispersive  
coupling $(\hbar g^2/\Delta) \sigma_z a^\dagger a$, this constitutes a
quantum non-demolition measurement of the qubit, which has already been
implemented experimentally \cite{Grajcar2004a, Wallraff2004a,
Johansson2006a, Sillanpaa2006a}.  
In turn, the qubit energy splitting is shifted depending on the mean
photon number $n = \langle a^\dagger a \rangle$.  Accordingly, one can
also measure the mean photon number, and even perform a full
quantum state tomography of the oscillator state \cite{Neeley2008a}.
Note also that
besides the condition of $\lambda$ being small, the perturbational result
\eqref{HJCdisp} is accurate only if the mean photon number $\langle
n\rangle$ does not exceed the critical value $n_\mathrm{crit} = 1/4
\lambda^2$. For larger photon numbers, higher powers of the number
operator $a^\dagger a$ must be taken into account
\cite{Boissonneault2009a, Boissonneault2009b} .   
Henceforth, we restrict ourselves to the so-called linear dispersive
regime, in which the photon number is clearly below the critical value
$n_\mathrm{crit}$. 


\section{Dispersive regime beyond RWA}
\label{sec:dispn}

It is now our aim to treat the original Hamiltonian \eqref{HRabi} in the
dispersive limit accordingly, i.e., to derive an expression that
corresponds to Eq.~\eqref{HJCdisp} but is valid in the full dispersive
regime defined by inequality \eqref{dispersiveNonRWA}.  Going
beyond RWA, we have to keep the counter-rotating coupling terms
\begin{equation}
 Y_\pm = \sigma^+a^\dagger \pm \sigma^-a \;,
\end{equation}
which are relevant if either of the relations $g \ll
\min \{ \epsilon, \omega\}$ or $|\epsilon-\omega| \ll \epsilon+\omega$
is violated. 
Separating again the qubit-oscillator coupling from the bare terms, we 
rewrite Hamiltonian \eqref{HRabi} as
\begin{eqnarray}
\label{HRabicompact}
 H = H_0 + \hbar g X_+ + \hbar g Y_+ \;,
\end{eqnarray}
which differs from $H_\mathrm{RWA}$ by the last term.
It will turn out that a unitary transformation corresponding to
Eq.~\eqref{transfone} is achieved by the operator
\begin{equation}
\label{transfone1}
D = e^{\lambda X_- + \olamb Y_-} \;.
\end{equation}
Here we have introduced the parameter
\begin{equation}
 \olamb =  \frac{g}{\epsilon+\omega} =  \frac{g}{\nu} \; ,
\end{equation}
which obviously fulfills the relation $\olamb < \lambda$, since 
$\epsilon$ and $\omega$ are positive.  Thus, whenever $\lambda$ is
small, $\olamb$ is small as well.  Nevertheless, under condition
\eqref{dispersiveNonRWA}, $\lambda$ and $\olamb$ may be of the same
order. 

Proceeding as in Sec.~\ref{sec:disp-JC}, we define the dispersive
Hamiltonian $H_\mathrm{disp} = D^\dagger H D$.  Using the commutation
relations $[Y_+, Y_-] =\sigma^ z ( 2 a^\dagger a + 1)  -1$,
$[\epsilon/2 \,\sigma^ z + \omega a^\dagger a , Y_-] 
= - (\epsilon + \omega ) Y_+ $ and $[Y_\pm, X_\mp ]= \sigma^ z \{ a^2 +
(a^\dagger)^2 \}$, we obtain the
expression 
\begin{equation}
\label{HRabidisp}
\begin{split}
 H_{\rm disp}
={} & 
\frac{\hbar\epsilon}{2} \sigma^ z
+ \hbar \omega a^\dagger a
\\ &+
\frac{\hbar g^2}{2} \left( \frac{1}{\Delta} + \frac{1}{\nu} \right)
   \sigma^z (a^\dagger +  a)^2   \;,
\end{split}
\end{equation}
which is valid up to second order in the dimensionless coupling
parameters $\lambda$ and $\olamb$.  

As compared to the RWA result \eqref{HJCdisp}, we find two
differences: First, the prefactor of the coupling has a
contribution that obviously stems from
$\olamb$.  Second and more importantly, the coupling is no longer
proportional to the number operator $a^\dagger a$, but rather to
$(a^\dagger + a)^2$.  Thus, the operator $Y_\pm$ has turned into
the counter-rotating contributions $a^2$ and $(a^\dagger)^2$.
For this reason, the dispersive Hamiltonian \eqref{HRabidisp} is not
diagonal in the eigenbasis of the uncoupled Hamiltonian $H_0$.

Nevertheless, it is possible to interpret the result as a qubit-state 
dependent frequency shift by the following reasoning.
Let us interpret $\hbar\omega a^\dagger a$ as the Hamiltonian of a
particle with unit mass in the potential $\frac{1}{2}\omega^2 x^2$,
where $x = \sqrt{\hbar/2\omega}(a^\dagger + a)$.  Then the
qubit-oscillator coupling in Eq.~\eqref{HRabidisp} modifies the
potential curvature $\omega^2$, such that the oscillator frequency
undergoes a shift according to
\begin{equation}
\omega \to \overline \omega = \omega \sqrt {1 \pm
  \frac{2g^2}{\omega} \left( \frac{1}{\Delta} +
    \frac{1}{\nu}\right)} . \label{eq:freqtrafo}
\end{equation}
Again the sign depends on the qubit state.  To be consistent with the
second-order approximation in $g$, we have to expand also the square
root to that order. This complies with the experimentally interesting
parameter regime where $g <\omega$.  We finally obtain
\begin{equation}
 \label{wrnonRWA}
 \overline \omega = \omega \pm g^2 \left ( \frac{1}{\Delta} + \frac{1}{\nu}
 \right ) .
\end{equation}
As for the RWA Hamiltonian, we find that the qubit state shifts the
resonance frequency of the oscillator.  This result is not only of
appealing simplicity, but also has a rather important consequence:
Dispersive readout is possible even when the qubit-oscillator coupling
is so strong that condition \eqref{dispersiveRWA} cannot be fulfilled,
that is, when the RWA result is not valid.

\begin{figure}
 \includegraphics[width=\columnwidth]{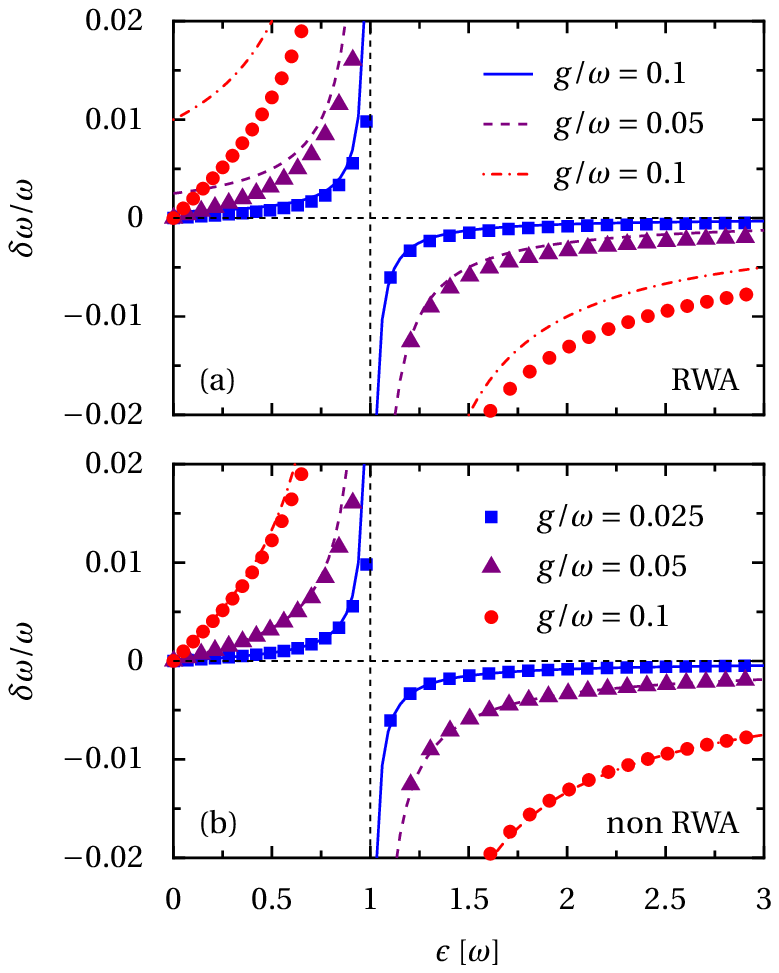}
\caption{Oscillator frequency shift as function or the qubit splitting
  $\epsilon=\omega+\Delta$ for the spin state
$|{\downarrow}\rangle$ obtained (a) within RWA, Eq.~\eqref{shiftRWA},
and (b) beyond RWA, Eq.~\eqref{wrnonRWA}.  The lines mark the
analytical results, while the symbols refer to the numerically
obtained splitting between the ground state and the first excited
state in the subspace of the qubit state $|{\downarrow}\rangle$.
}
\label{fig:wr} 
\end{figure}

For a quantitative analysis of our analytical findings, we compare the 
frequency shifts \eqref{shiftRWA} and \eqref{wrnonRWA} with numerical
results.  In doing so, we diagonalize the Hamiltonian \eqref{HRabi} in
the subspace of the qubit state $|{\downarrow}\rangle$, where
$\sigma^z |{\downarrow}\rangle = -|{\downarrow}\rangle$.  The results
are depicted in Fig.~\ref{fig:wr}.
%

For a qubit splitting $\epsilon$ close to the cavity frequency
$\omega$, i.e., outside the dispersive regime, the analytically
obtained frequency shifts diverge.  This behavior is certainly
expected for an expansion in $g/\Delta$.  For a relatively small
coupling $g/\omega\,{\lesssim}\, 0.025$, the RWA result [panel (a)]
agrees  very well with the numerical data in the dispersive regime.
In the case of larger coupling strengths, $g/\omega\,{\gtrsim}\,
0.05$, the predictions from RWA exhibit clear differences.  The
general tendency is that RWA overestimates the frequency shift for
blue detuning $\Delta \,{=}\, \epsilon-\omega<0$, while it predicts a too
small shift for red detuning.

The data shown in panel (b) demonstrates that the treatment beyond RWA
yields the correct frequency shift in the entire dispersive regime,
i.e., whenever the detuning significantly exceeds the
coupling, $|\Delta|\gg g$.  Thus, as long as the coupling remains much
smaller than the oscillator frequency, $g\ll\omega$, it is always
possible to tune the qubit splitting $\epsilon$ into a regime in which 
\ref{dispersiveNonRWA} is fulfilled.  Moreover, the excellent
quantitative agreement 
of our analytical result \eqref{wrnonRWA} with the numerically exact
solution indicates the feasibility to determine $g$ from measurements
in the strong-coupling limit \cite{Bourassa2009a}.   

A particular limit is $\Delta\to -\omega$, which corresponds to a
vanishing qubit splitting, $\epsilon\to 0$.  In this case it is
obvious from Hamiltonian \eqref{HRabi} that the coupling to the 
qubit merely entails a linear displacement of the oscillator
coordinate, while the oscillator frequency remains unaffected.  This
limit is perfectly reproduced by our non-RWA result \eqref{wrnonRWA},
irrespective of the coupling strength.  The RWA result, by contrast,
predicts a spurious frequency shift, indicating the failure of RWA.


\section{Several qubits in a cavity}
\label{sec:multi-qb}

An experimentally relevant generalization of the model \eqref{HRabi}
is the case of several qubits coupling to the same oscillator.  The
corresponding Hamiltonian reads \cite{Blais2007a} 
\begin{equation}
\label{HTCn}
 H = \hbar \sum_j \frac{\epsilon_j}{2} \sigma^z_j + \hbar \omega
 a^\dagger a +  \hbar \sum_j g_j^{\vphantom{\dagger}} \sigma^x_j
 (a^\dagger +a) , 
\end{equation}
where the index $j$ labels the qubits.  As for the one-qubit case, the 
rotating wave-approximation is frequently applied and yields the 
Tavis-Cummings Hamiltonian \cite{Tavis1968a}
\begin{equation}
\label{HTC}
 H = \hbar \sum \frac{\epsilon_j}{2} \sigma^z_j + \hbar \omega
 a^\dagger a +  \hbar \sum g_j X^j_+ \;,
\end{equation}
where  $X_\pm^j = \sigma^-_j a^\dagger_j \pm \sigma^+_j
a_j^{\vphantom{\dagger}} $, cf.\ Eq.~\eqref{H0Xpm}.

%
\subsection {Dispersive theory within RWA}

We obtain for each qubit the dimensionless coupling parameter 
$\lambda_j = g_j/(\epsilon_j - \omega)$. The dispersive limit is now 
determined by $|\lambda_j|\ll 1$ for all $j$. Effective decoupling of the
qubits and the cavity to second order is then achieved via a 
transformation with the unitary operator
$ \exp({-\sum_j \lambda_j X_-^j})$, cf.\ Eq.~\eqref{transfone}. The
resulting dispersive Hamiltonian reads \cite{Blais2007a}
\begin{equation}
\label{HTCdisp}
\begin{split}
 H_{\rm disp}
={}& \hbar\omega a^\dagger a + \frac{\hbar}{2} \sum_j
     \left(\epsilon_j+ \frac{g^2}{\Delta_j}\right) \sigma^z_i 
\\
&+ \sum \frac{g^2}{\Delta_j} a^\dagger a \sigma_z^j
 + \sum_{j>k} J_{jk} (\sigma^-_j \sigma_k^+ + \sigma^+_j \sigma_k^-) .
\end{split}
\end{equation}
Remarkably, the oscillator entails an effective coupling between the
qubits with the strength
\begin{equation}
\label{JRWA}
 J_{jk}= g_j g_k \left( \frac{1}{ \Delta_j } + \frac{1}{ \Delta_k }
 \right), 
\end{equation}
which has already been observed experimentally \cite{Majer2007a}.  It 
has been proposed to employ this interaction for building qubit
networks \cite{Helmer2009a} and for generating qubit-qubit entanglement
\cite{Helmer2009b, Hutchison2009a, Bishop2009a}. Moreover, quantum
tomography of a two-qubit state has been implemented by probing the
cavity at its bare resonance frequency \cite{Flipp2009a}. In this
scenario  the oscillator frequency exhibits a shift depending on a
collective coordinate of all qubits. Consequently, the cavity response
experiences a phase shift from the ingoing signal, which in turn
contains information about that collective qubit coordinate.

\subsection{Dispersive theory beyond RWA}

As in Sec.~\ref{sec:dispn} for the one-qubit case, we now extend the
dispersive theory of the Tavis-Cummings model beyond RWA, taking into
account the counter-rotating terms of the Hamiltonian (\ref{HTCn}). 
In analogy to transformation~(\ref{transfone1}), we employ the ansatz 
\begin{equation}\label{eq:transf-taviscum}
 D= e^{\sum \lambda_j X_-^j + \olamb_j Y_-^j}  ,
\end{equation}
where
$Y_-^j = \sigma^- _j a - \sigma^+_j a^\dagger$
and $\olamb_j = g/(\nu_j)$. 
Following the lines of Sec.~\ref{sec:dispn}, i.e.\ expanding the
transformed Hamiltonian to second order in $\lambda$ and $\olamb$,
we obtain the dispersive Hamiltonian
\begin{eqnarray}
\nonumber H_{\rm disp}
&=& D^\dagger H D \\
&=& \hbar\omega a^\dagger a 
  + \frac{\hbar}{2} \sum_j \epsilon_j  \sigma^z_i 
\nonumber\\
&& +
\frac{1}{2}\sum_j g_j^2 \left (\frac{1}{\Delta_j}+ \frac{1}{\nu_j}
\right ) ( a^\dagger + a)^2 \sigma^z_j  
\nonumber\\
&&+ \sum_{j>k} \overline J_{jk} \sigma^x_j \sigma_k^x \, .
\label{HTCndisp}
\end{eqnarray}
%
We have introduced the modified coupling strength 
\begin{eqnarray}
\label{Jn}
 \overline{J}_{jk}= 
g_i g_k
\left (
\frac{1}{ \Delta_j } + \frac{1}{ \Delta_k } 
- \frac{1}{ \nu_j } - \frac{1}{ \nu_k } 
\right ) ,
\end{eqnarray}
which describes the effective interaction between qubits $j$ and $k$,
and represents the extension of Eq.~(\ref{JRWA}) beyond RWA.  The
dispersive shifts of the qubit and cavity frequencies, given by the
second and third term of Eq.~(\ref{HTCndisp}), are equally modified as 
compared to the RWA result (\ref{HTCdisp}). 
\void{The same holds true for the effective qubit-qubit interactions
  $\overline{J}_{jk}$.}

Interestingly enough, the effective qubit-qubit interaction in
Eq.~(\ref{HTCndisp}) is of the Ising type $\sigma^x_j \sigma^x _k$,
whereas RWA predicts the isotropic XY interaction $\sigma^+ _j \sigma^-
_k +\sigma^- _j \sigma^+ _k$, see Eq.~(\ref{HTCdisp}).  Thus, the
treatment beyond RWA predicts a qualitatively different effective
model and not merely a renormalization of parameters. The Ising term
even persists in the limit $1/\Delta_j \gg
1/(\nu_j)$. Nevertheless, one can recover the RWA Hamiltonian
(\ref{HTCdisp}) by writing the interaction term as $\sigma^x_j \sigma^x_k
= \sigma_j^+ \sigma_k^- + \sigma_j^+ \sigma_k^+ + {\rm h.c.}$ and
performing a RWA for the Ising coupling.  This corresponds to
discarding small-weighted, rapidly oscillating terms of the type
$\overline J_{jk} \sigma_j^+ \sigma_k^+ +  {\rm h.c.}$

The difference between the effective models \eqref{HTCdisp} and
\eqref{HTCndisp} has some physically relevant consequences. First, in
contrast to the RWA result \eqref{HTCdisp}, Hamiltonian
(\ref{HTCndisp}) does not conserve the number of qubit excitations,
which will affect the design of two-qubit gates \cite{Blais2007a}.
Moreover, both models possess different spectra, which influences
entanglement creation.  For instance, the ground state of the
Hamiltonian (\ref{HTCndisp}) for two degenerate qubits
($\epsilon_1\,{=}\,\epsilon_2$) coupled to one cavity is $|0\rangle
[|{\down \down}\rangle \,{-}\, (J/2\epsilon) |{\up \up} \rangle ]$ and thus 
exhibits qubit-qubit entanglement.  By contrast, the corresponding
ground state of the multi-qubit RWA Hamiltonian (\ref{HTCdisp}) is the
product state $|0\rangle |{\down \down}\rangle$.
For the case of the Hamiltonian (\ref{HTCndisp}), thermal
qubit-qubit entanglement will consequently be present at zero
temperature and even at thermal equilibrium \cite{Amico2008,
  Lagmago2002a}. 


\section{Summary}

We have generalized the dispersive theory for a qubit coupled to a
harmonic oscillator to the case of far detuning.  In this limit, it is
no longer possible to treat the qubit-oscillator interaction Hamiltonian
within the rotating-wave approximation. Therefore, previous derivations
need some refinement.  It has turned out that diagonalizing
the Hamiltonian analytically up to second order in the coupling
constant is possible as well beyond RWA.  The central result is that
as within RWA, the oscillator experiences a shift of its resonance
frequency, the sign of the shift depending on the qubit state.  In  
this respect, the difference between both approaches seems to be
merely quantitative. 
Nevertheless, our result implies an important fact for currently
devised qubit-oscillator experiments with ultra-strong cavity-qubit
coupling: Dispersive qubit readout is possible as well in that regime.
The comparison with numerical results has confirmed that our approach
is quantitatively satisfactory in the whole dispersive regime.

The corresponding treatment of many qubits coupled to the same
oscillator is equally possible.  In such architectures, the oscillator
mediates an effective qubit-qubit interaction which may be used for
gate operations and entanglement creation.  We have revealed that the
form of the effective interaction depends on whether or not one
employs RWA.  While RWA predicts an isotropic XY interaction, the 
inclusion of the counter-rotating terms yields an interaction of 
Ising type. This difference impacts on various proposed entanglement 
creation protocols as soon as they operate in the far-detuned
dispersive regime.


\begin{acknowledgments}
We would like to thank Michele Campisi, Frank Deppe, Johannes Hausinger, Matteo Mariantoni and  Enrique Solano for discussions. 
We gratefully acknowledge financial support by the German Excellence
Initiative via the ``Nanosystems Initiative Munich (NIM)''.
This work has been supported by DFG through SFB 484 and SFB 631.
\end{acknowledgments}



\end{document}